\documentclass[epj]{svjour}
\usepackage{epsfig}
\usepackage{amsmath}

\def\bsigma{\mbox{\boldmath$\sigma$}}

\def\e{\kern+.6ex\lower.42ex\hbox{$\scriptstyle \iota$}\kern-1.20ex e}
\def\a{\kern+.6ex\lower.42ex\hbox{$\scriptstyle \iota$}\kern-1.20ex a}
\def\e{\kern+.5ex\lower.42ex\hbox{$\scriptstyle \iota$}\kern-1.10ex e}

\newcommand{\fet}[1]{\mbox{\boldmath $ #1 $}} 

\begin{document}
\title{
A new way to perform partial wave decompositions of few-nucleon forces
}
 
\author{J.~Golak\inst{1} \and
               D.~Rozp{\e}dzik\inst{1} \and
               R.~Skibi\'nski\inst{1} \and
               K.~Topolnicki\inst{1} \and
               H.~Wita{\l}a\inst{1} \and
               W. Gl\"ockle\inst{2} \and
               A. Nogga\inst{{3,4}} \and
               E. Epelbaum\inst{{3,5}} \and
               H. Kamada\inst{6} \and
               Ch. Elster\inst{7} \and
               I. Fachruddin\inst{8} }
\institute{M. Smoluchowski Institute of Physics, Jagiellonian
University, PL-30059 Krak\'ow, Poland
\and 
Institut f\"ur Theoretische Physik II,
Ruhr-Universit\"at Bochum, D-44780 Bochum, Germany
\and 
Forschungszentrum J\"ulich, 
          Institut f\"ur Kernphysik (Theorie)
          and  J\"ulich Center for Hadron Physics, D-52425 J\"ulich, Germany
\and 
Forschungszentrum J\"ulich, 
                Institute for Advanced Simulation, 
                D-52425 J\"ulich, Germany
\and           
Helmholtz-Institut f\"ur Strahlen- und Kernphysik (Theorie) 
and Bethe Center for Theoretical Physics, Universit\"at Bonn, 
                  D-53115 Bonn, Germany
\and 
Department of Physics, Faculty of Engineering, 
Kyushu Institute of Technology, Kitakyushu 804-8550, Japan 
\and 
Institute of Nuclear and Particle Physics, 
Department of Physics and Astronomy, Ohio University, Athens, OH 45701, USA \and 
Departemen Fisika, Universitas Indonesia, 
Depok 16424, Indonesia}

\date{Received: date / Revised version: date}
\abstract{
We formulate  a general and exact method 
of partial wave decomposition (PWD) of any nucleon-nucleon (NN) potential
and any three-nucleon (3N) force. The approach allows one to efficiently use symbolic algebra software 
to generate the interaction dependent part of the program code 
calculating the interaction. We demonstrate the feasibility
of this approach for 
the one-boson exchange BonnB potential, a 
recent nucleon-nucleon chiral force and the chiral two-pion-exchange 
three-nucleon force. In all cases  very good agreement between 
the new and the traditional PWD is found.
The automated PWD offered by the new approach is of the utmost 
importance in view of future applications of numerous chiral N3LO 
contributions to the 3N force in three nucleon calculations.
\PACS{
      {21.45.-v}{few-body systems}   \and
      {21.30.-x}{nuclear forces} \and 
      {21.45.Bc}{two-nucleon system} \and 
      {21.45.Ff}{three-nucleon forces}
     } 
} 
\maketitle

\section{Introduction}
\label{I}

The standard way to set up calculations of two- and three-nucleon systems 
is a partial wave decomposition (PWD). Especially at low energies, i.e. below
the pion production threshold, this procedure is still most commonly used,
despite the advent of the approaches which use a direct three-dimensional 
notation \cite{charl1,hang,imam,teheran}. 

Recently we proposed a formulation of the two- and three-nucleon
system~\cite{2N3N,new2N}, which is based on
scalar spin-momentum operators
and accompanying scalar functions depending only on the momenta of the
system. 
This formulation is based on the most general operator structure 
a nuclear force given in momentum space can have. 
The two- and three-nucleon equations are obtained by carrying out
traces over the spin-momentum operators building the nuclear force.
The same approach can be used to obtain partial wave projected
matrix elements of the potential and the transition operators.
Taking traces of spin-momentum operators lends itself to the use of
symbolic algebra software to obtain general expressions
for those matrix elements.

In this work we demonstrate that by algebraic operations 
general expressions
for the partial wave decomposition of any nucleon-nucleon (NN) potential 
can be obtained.
This method will be presented in Sec.~\ref{II}, and explicit 
expressions for calculating specific matrix elements of NN potentials
are given in an appendix. 
Numerical comparisons between our suggested methods and the
standards partial wave decomposition  using recent chiral NN forces
\cite{evgeny.report} and the one-boson-exchange NN potential BonnB ~\cite{machl} 
are presented in Sec.~\ref{III}.
In Sec.~\ref{IV} we demonstrate that the same method can be extended 
to the PWD of a three-nucleon 
force. As a numerical example given in Sec.~\ref{V} we take 
the two-pion-exchange (TPE) chiral NNLO three-nucleon force.
Finally we conclude in Sec.~\ref{VI}.

\section{Partial wave decomposition of the NN potential}
\label{II}

We start, as in Ref.~\cite{new2N}, by projecting the NN potential on 
the two-nucleon (2N)
isospin states $ | (\frac12 \frac12 ) t m_t \rangle \equiv | t m_t \rangle$.
Furthermore, we assume that there is no isospin mixing but 
allow for charge independence 
and charge symmetry breaking, and thus for a dependence on $m_t$:
\begin{eqnarray}
\langle  t' m_t' | V | t m_t \rangle  =   \delta_{ t' t} \delta_{ m_t'
m_t } V^ { t m_t}.
\label{eq:2.1}
\end{eqnarray}
It is well known that the most general form of the NN force, which is
invariant under rotations, parity and time reversal can be expressed
by six scalar spin-momentum operators \cite{wolfenstein}, which we choose as
\begin{eqnarray}
w_1 ( {\bsigma}(1),{\bsigma}(2), {\fet p'}, {\fet p})&  = &  1 \ , \cr
w_2 ( {\bsigma}(1),{\bsigma}(2), {\fet p'}, {\fet p})&  = & {\bsigma}(1) \cdot {\bsigma}(2)  \ , \cr
w_3 ( {\bsigma}(1),{\bsigma}(2), {\fet p'}, {\fet p)}&  = & i \; ( {\bsigma}(1)
+ {\bsigma}(2) ) \cdot ( {\fet p} \times {\fet p'})  \ , \cr
w_4 ( {\bsigma}(1),{\bsigma}(2), {\fet p'}, {\fet p})&  = & {\bsigma}(1)
\cdot ( {\fet p} \times {\fet p'}) \; {\bsigma}(2) \cdot ( {\fet p} \times {\fet p'}) \ , \cr
w_5 ( {\bsigma}(1),{\bsigma}(2), {\fet p'}, {\fet p})&  = & {\bsigma}(1)
\cdot  ({\fet p'} + {\fet p}) \; {\bsigma}(2) \cdot  ({\fet p'} + {\fet p}) \ , \cr
w_6 ( {\bsigma}(1),{\bsigma}(2), {\fet p'}, {\fet p})&  = & {\bsigma}(1)
\cdot ( {\fet p'} - {\fet p}) \; {\bsigma}(2) \cdot  ( {\fet p'} - {\fet p})  \ . \cr
& & 
\label{eq:2.2}
\end{eqnarray}
Thus the isospin projected potential can be expresses as
\begin{eqnarray}
V^ { t m_t} (  {\fet p'}, {\fet p}) = \sum\limits_{i=1}^6 f_i (  {\fet p'}, {\fet p}) 
w_i ( {\bsigma}(1),{\bsigma}(2), {\fet p'}, {\fet p}) .
\label{decomposition}
\end{eqnarray}
The expansion coefficient here are 
scalar functions $f_i (  {\fet p'}, {\fet p})$ that depend on two vector
momenta ${\fet p'}$ and ${\fet p}$, more specifically
 on the magnitudes of the vectors and 
and the cosine of the relative angle between them.
In order to determine the functions $f_i$, we evaluate the spin
dependence analytically by taking traces with the operators of
Eq.~(\ref{eq:2.2}) and thus arrive at 
a system of six coupled linear equations 
\begin{eqnarray}
\sum\limits_{j=1}^6 {\rm Tr} \left( w_i \, w_j \right) \, f_i = 
{\rm Tr} \left(  V^ {t m_t} w_i \right), \ \ \ i=1,2, \dots, 6 ,
\label{traces}
\end{eqnarray}
which has a unique solution provided that $p' \ne p$.
(When $p' = p$ only five out of the six operators $w_i$ 
are sufficient, since $w_2$ is linearly dependent on $w_4$, $w_5$, and
$w_6$.)

Our task is to obtain matrix elements of $V^ { t m_t} $ in the basis of states 
$ \mid p (l s ) j m_j \rangle $, where the relative angular momentum $l$
and the total spin $s$ are coupled to the
 total angular momentum $j$ with its
projection $m_j$.
When calculating  NN observables one usually sums angular momenta $j$ up
to a certain $j_{max}$ at which the calculation is converged. 
For calculations of NN observables below the pion production threshold
convergence is reached for $j_{max} \leq 8$.

To obtain the potential matrix element in the basis $ \mid p (l s ) j
m_j \rangle $, the four-fold integral
\begin{eqnarray}
\lefteqn{\langle p'(l's ) j m_j \mid V^{t m_t}\mid p (ls) j m_j
\rangle = } \cr
& &= \int d \hat{p\,}'  \int d \hat {p} \; 
 \sum\limits_{m_l'} c ( l',s,j;m_l', m_j - m_l',m_j)  \; \cr 
& & \quad \times \sum\limits_{m_l} c ( l,s,j;m_l, m_j - m_l,m_j) 
 \ Y^*_{l' \, m_l'} ( \theta', \phi' )  \;
Y_{l \, m_l} ( \theta, \phi )  \; \cr 
& & \quad \times  \langle s \, m_j - m_l' \mid V^ { t m_t} ( {\fet p'} , {\fet p} ) \mid s  \,m_j - m_l \rangle 
\label{PWD1}
\end{eqnarray}
needs to be evaluated.
Here $ c ( l,s,j;m_l, m_j - m_l,m_j)$
are the standard Clebsch-Gordan coefficients,
and $ Y_{l \, m_l} ( \theta, \phi ) $ 
the spherical harmonics calculated for the angles corresponding to 
the directions of the momenta ${\fet p'}$ and ${\fet p}$.
This quantity does not actually depend on $m_j$, so instead of
Eq.~(\ref{PWD1}) we can calculate
\pagebreak
\begin{eqnarray}
H ( l',l,s,j ) \equiv 
\frac{1}{2 j +1} \sum\limits_{m_j=-j}^j \, 
\langle p' (l' s ) j m_j \mid V \mid p (l s ) j m_j \rangle  .\cr 
& & 
\end{eqnarray}
Since now the integrand is a scalar, it is possible to reduce 
the number of integrals to one:
\begin{eqnarray}
\lefteqn{H ( l',l,s,j )  = 
8 \pi^2 \, \int\limits_{-1}^{1} d ( \cos \theta' ) \, 
\frac{1}{2 j +1} \sum\limits_{m_j=-j}^j } \cr  
& &\times \sum\limits_{m_l'=-l'}^{l'} c ( l',s,j;m_l', m_j - m_l',m_j) \, \cr 
& & \times \sum\limits_{m_l=-l}^{l} c ( l,s,j;m_l, m_j - m_l,m_j) \cr
& & \times \  Y_{l' \, m_l'} ( \theta', 0  )  \,
Y^*_{l \, m_l} ( 0 , 0 ) \, \cr 
& & \times \langle s \, m_j - m_l' \mid V({\fet p'} , {\fet p} ) \mid s  \,m_j - m_l 
\rangle  . 
\label{main}
\end{eqnarray}
In the NN system we can choose the z-axis to be
\begin{eqnarray}
{\fet p} = ( 0 , 0, p).
\end{eqnarray}
Then the direction of the vector ${\fet p'}$ is given as
\begin{eqnarray}
{\fet p'} = ( p' \sin \theta', 0 , p' \cos \theta' ) .
\end{eqnarray}
Most importantly however, 
the matrix element in the 2N spin space, 
\begin{eqnarray}
\lefteqn{\langle s \, m_j - m_l' \mid V( {\fet p'} , {\fet p} ) \mid s  \,m_j -
m_l \rangle = } \cr
& &\langle s \, m_j - m_l' \mid \sum\limits_{i=1}^6 f_i ( p', p, x )  \cr 
& & \qquad \qquad w_i ( {\bsigma}(1),{\bsigma}(2), {\fet p'}, {\fet p}) 
\mid s  \,m_j - m_l \rangle 
\end{eqnarray}
can be calculated analytically~\footnote{
For this calculation symbolic software like 
{\em Mathematica \copyright}~\cite{math} proves very useful.
Here it is particularly simple with the concept of the
Kronecker product which makes such matrix elements simple 
matrix elements in the  four-dimensional space.}.
The three sums in Eq.~(\ref{main}) over $m_j$, $m_l'$ and $m_l$
can be written out explicitly so the integrand over $x\equiv \cos \theta'$
can be prepared  {\em once} for all NN potentials that are represented
 in the form of Eq.~(\ref{decomposition}). 
The resulting expression  is given in terms of $p$, $p'$, 
$x$ and the expansion coefficients 
$f(i) \equiv f_i ( p', p, x )$. 
As example, evaluating Eq.~(\ref{main}) for the $^1S_0$ channel  
leads to
\begin{eqnarray}
 H(0,0,0,0)& =&  2 \pi \int\limits_{-1}^{1} dx \, 
 \Big( f(1) - 3 f(2) + f(4) p^2 p'^2 ( x^2-1) \cr
 & &   -f(5) \left( p^2+ p'^2 +2 pp'x\right) \cr 
& & - f(6) \left({p}^2 +p'^2 -2pp'x\right) \Big).
\end{eqnarray}
Further examples are collected in Appendix~\ref{A}.

\section{Application of the new method to the 2N potential}
\label{III}

\begin{figure*}[hbt]\centering
\epsfig{file=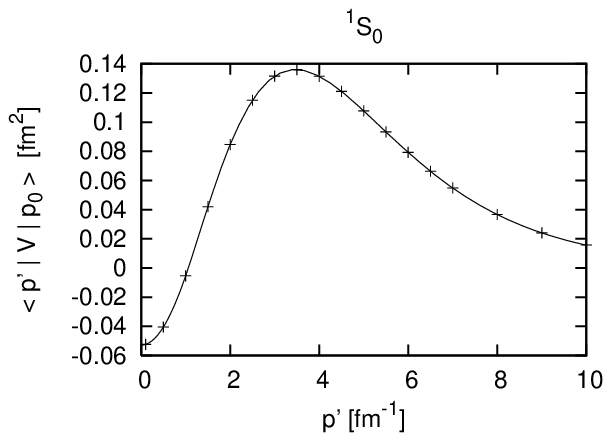,width=6.5cm,angle=0}
\epsfig{file=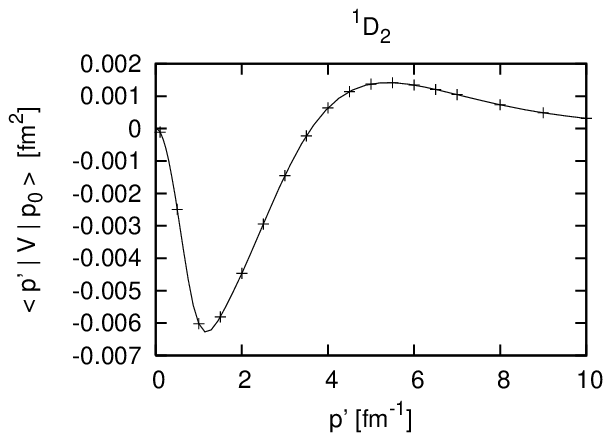,width=6.5cm,angle=0}
\epsfig{file=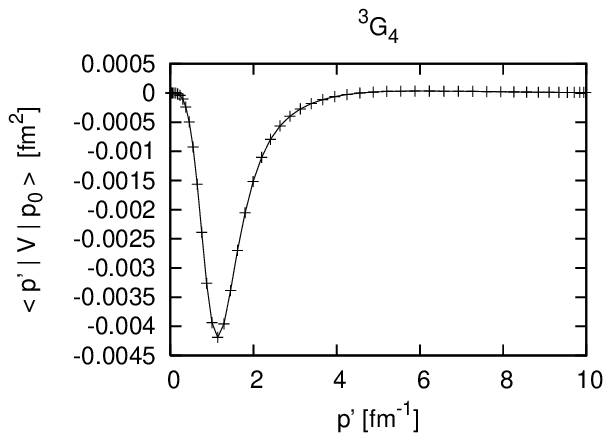,width=6.5cm,angle=0}
\epsfig{file=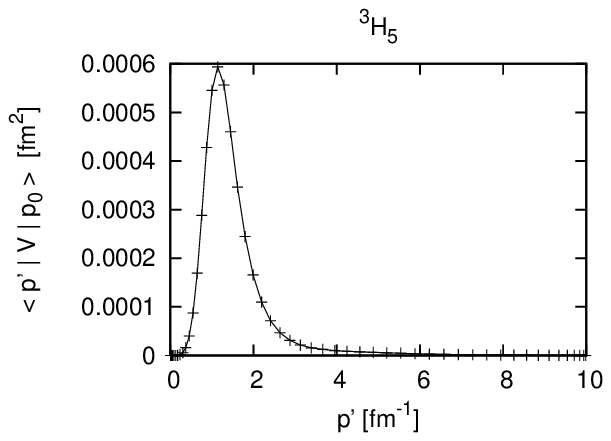,width=6.5cm,angle=0}
\caption{Comparison of the traditional (crosses) and the new (solid line) 
method of PWD for the Bonn B potential. The matrix elements of 
selected uncoupled channels with $t=1$ (${}^1S_0$, ${}^1D_2$) 
and $t=0$  (${}^3G_4$, ${}^3H_5$) are shown for a fixed 
value of $p_0$= 1 fm$^{-1}$ 
as a function of the $p'$ momentum.}
\label{f1}
\end{figure*}

\begin{figure*}[hp]\centering
\epsfig{file=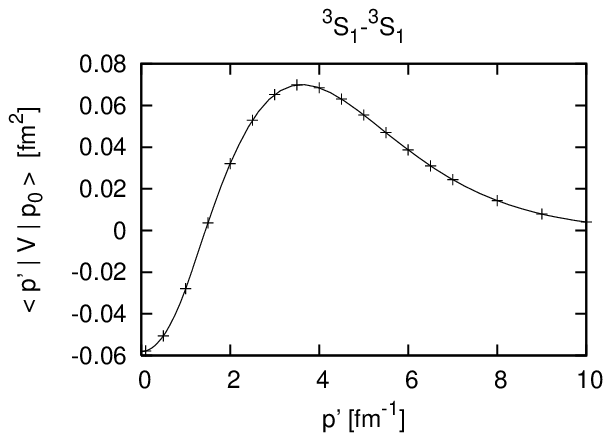,width=6.5cm,angle=0}
\epsfig{file=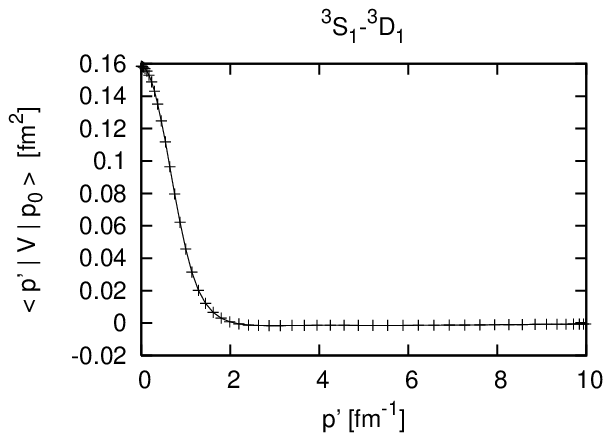,width=6.5cm,angle=0}
\epsfig{file=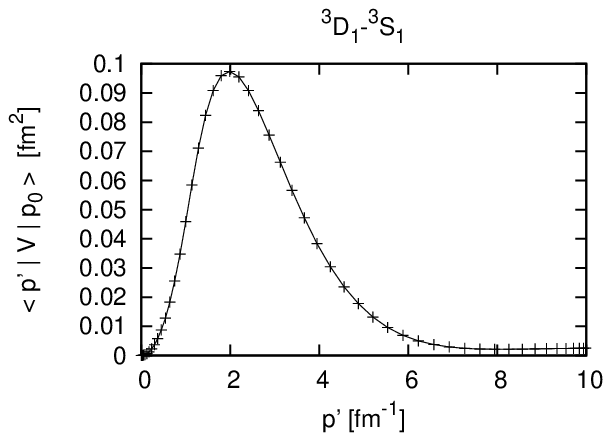,width=6.5cm,angle=0}
\epsfig{file=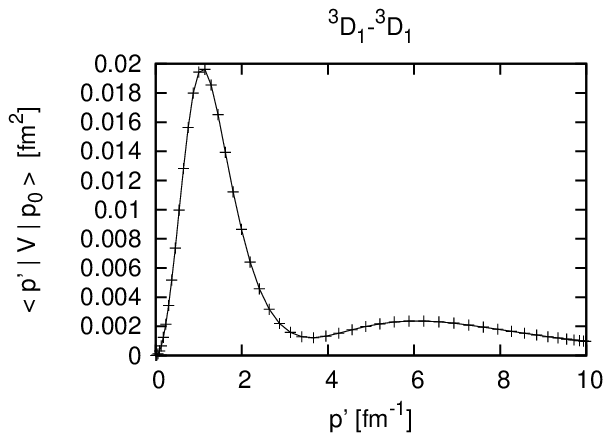,width=6.5cm,angle=0}
\caption{The same as in Fig.~\ref{f1} for one selected coupled channel case with $t=0$.}
\label{f2}
\end{figure*}

\begin{figure*}[hp]\centering
\epsfig{file=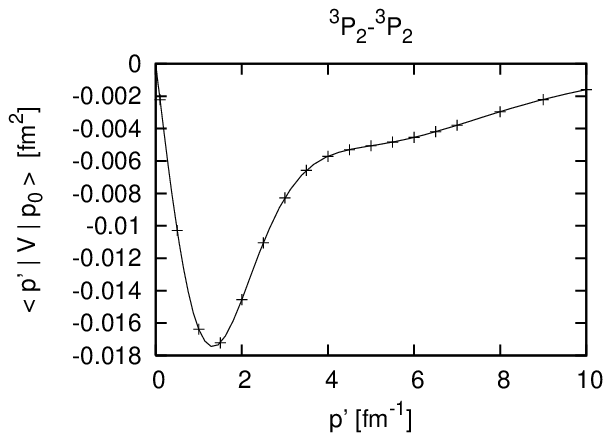,width=6.5cm,angle=0}
\epsfig{file=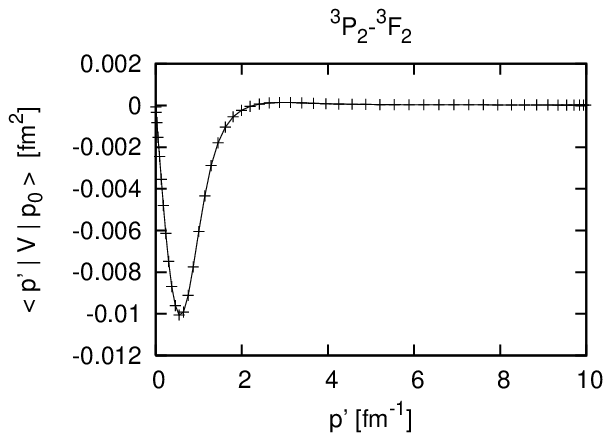,width=6.5cm,angle=0}
\epsfig{file=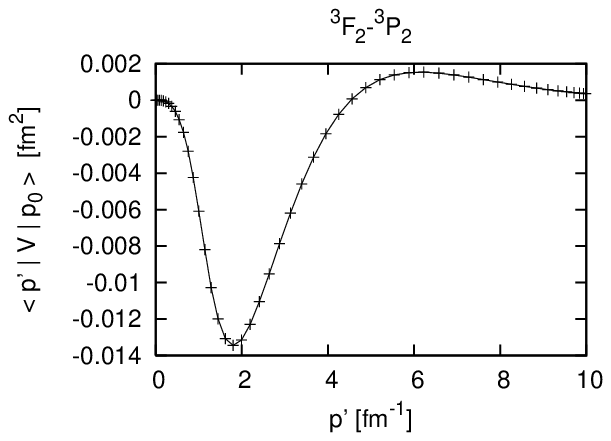,width=6.5cm,angle=0}
\epsfig{file=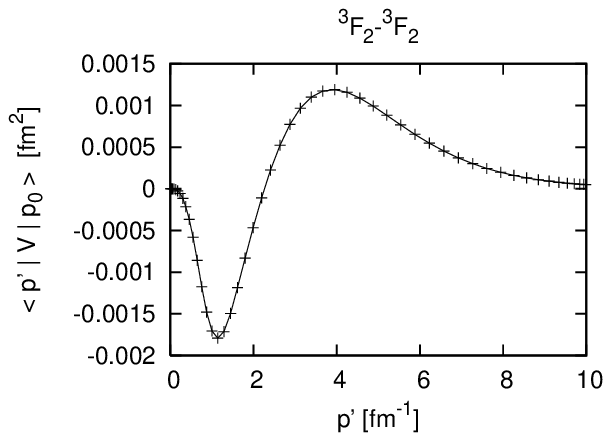,width=6.5cm,angle=0}
\caption{The same as in Fig.~\ref{f2} for one selected coupled channel case with $t=1$.}
\label{f3}
\end{figure*}

\begin{figure*}[hp]\centering
\epsfig{file=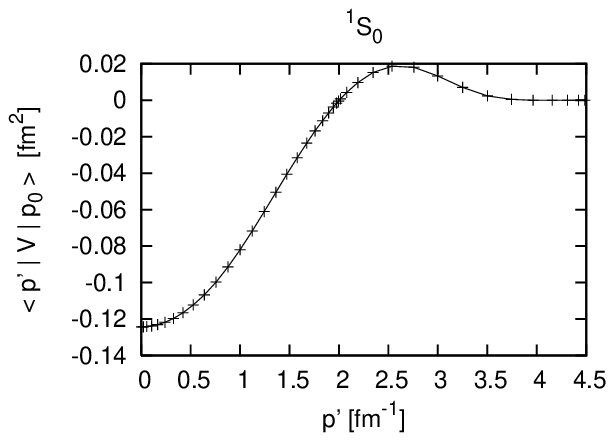,width=6.5cm,angle=0}
\epsfig{file=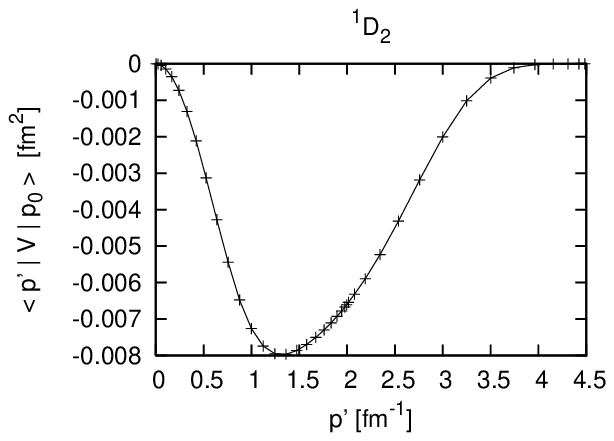,width=6.5cm,angle=0}
\epsfig{file=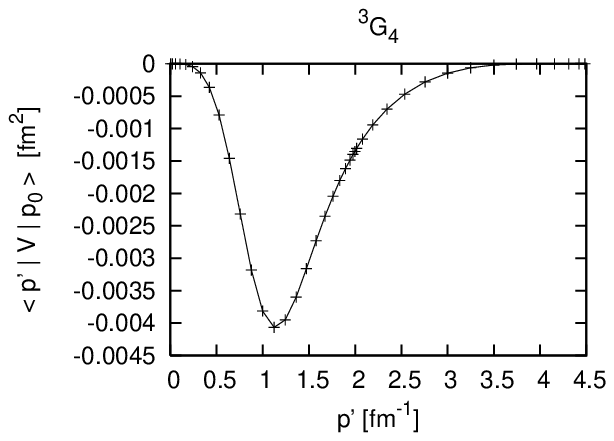,width=6.5cm,angle=0}
\epsfig{file=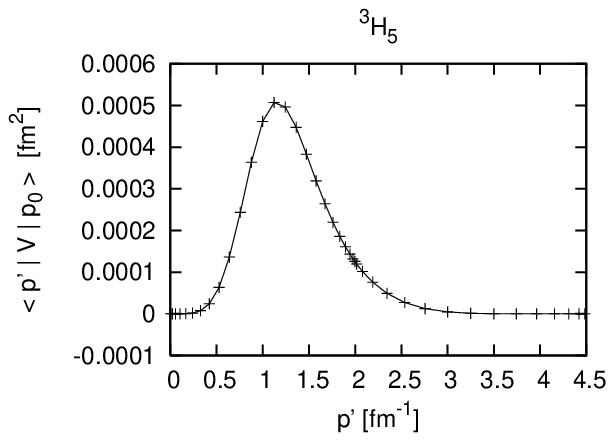,width=6.5cm,angle=0}
\caption{The same as in Fig.~\ref{f1} for the example of a chiral NNLO potential (see text for details).}
\label{f4}
\end{figure*}

\begin{figure*}[hp]\centering
\epsfig{file=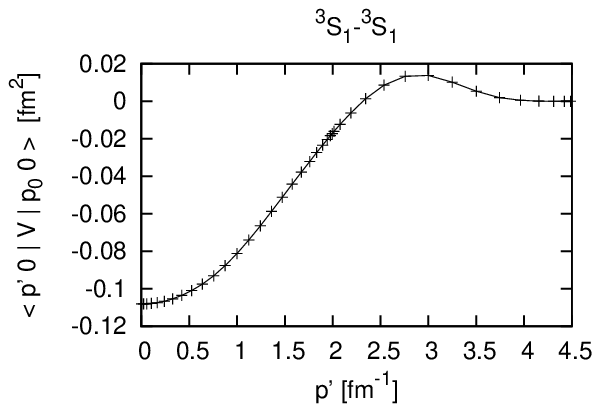,width=6.5cm,angle=0}
\epsfig{file=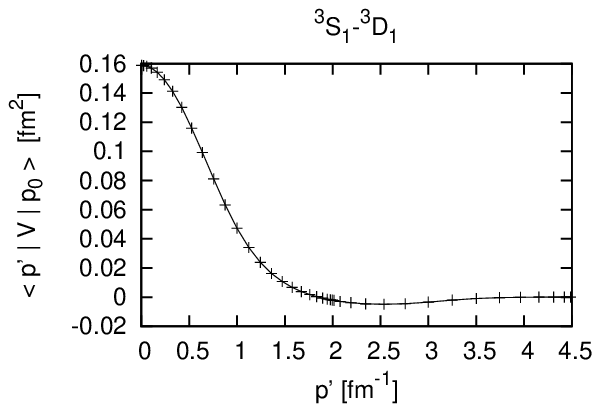,width=6.5cm,angle=0}
\epsfig{file=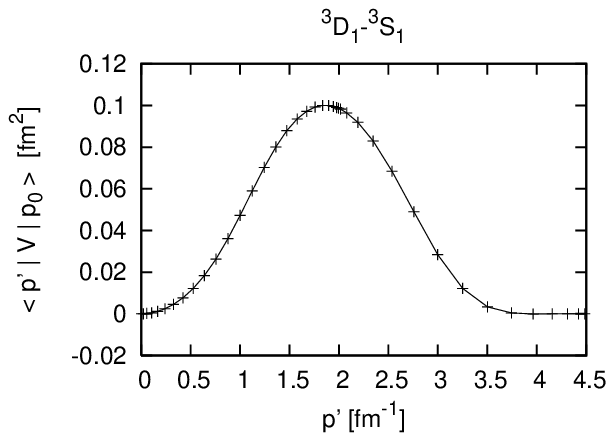,width=6.5cm,angle=0}
\epsfig{file=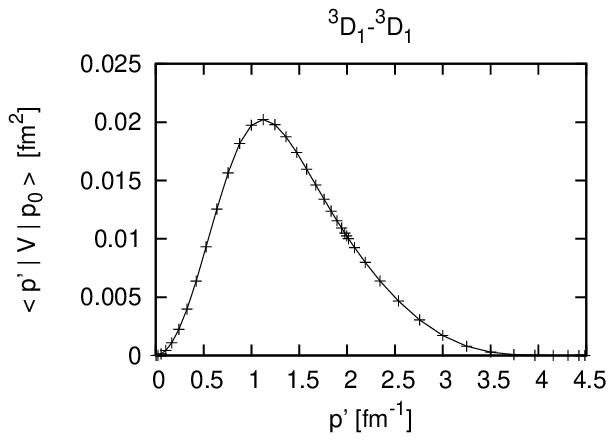,width=6.5cm,angle=0}
\caption{The same as in Fig.~\ref{f2} for the selected chiral NNLO potential.}
\label{f5}
\end{figure*}

\begin{figure*}[hp]\centering
\epsfig{file=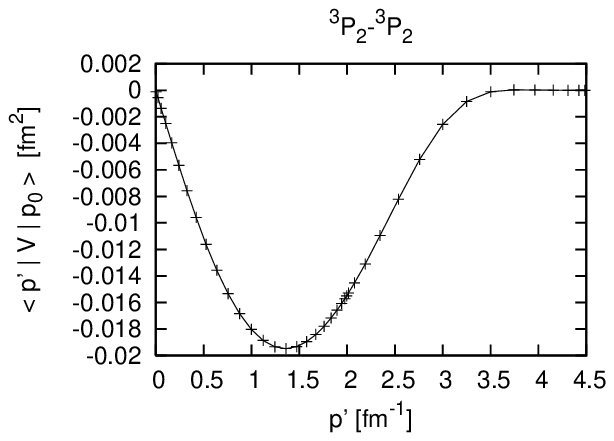,width=6.5cm,angle=0}
\epsfig{file=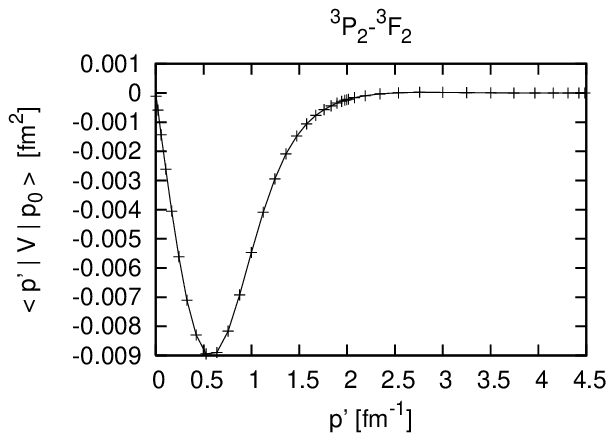,width=6.5cm,angle=0}
\epsfig{file=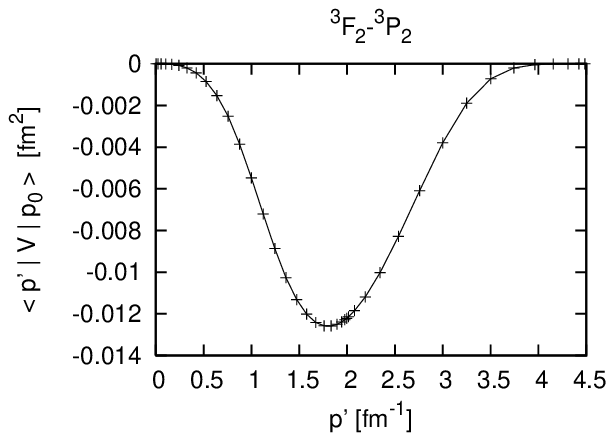,width=6.5cm,angle=0}
\epsfig{file=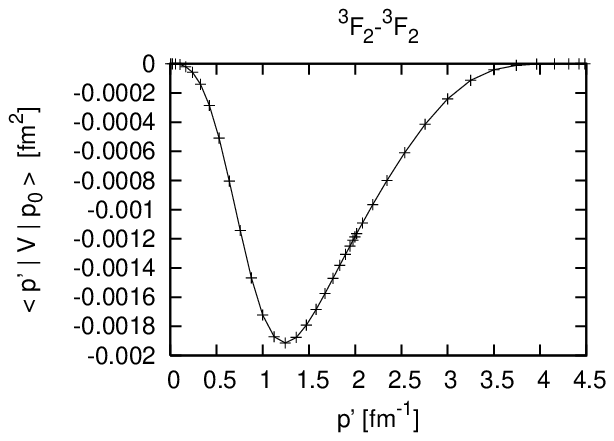,width=6.5cm,angle=0}
\caption{The same as in Fig.~\ref{f3} for the selected chiral NNLO potential.}
\label{f6}
\end{figure*}

In this section we want to give two examples a PWD 
of an NN potential based on our new method and compare with results
obtained in the traditional way.
First we consider the one-boson exchange potential  BonnB~\cite{machl}. 
Here we use the operator form of this potential presented in \cite{new2N}.
In Figs.~\ref{f1}--\ref{f6} we show matrix elements 
$\langle p' (l' s ) j m_j \mid V^ { t m_t}  \mid p_0 (l s ) j m_j \rangle $
for several partial waves for a fixed value of $p_0$~=~1~fm$^{-1}$
as a function of the momentum $p'$. 
In Figs.~1 through 3 selected partial wave projected potential amplitudes 
for the BonnB potential are shown. The agreement of the
matrix elements calculated  with our operator based
formulation and the original PWD based on
a helicity formulation~\cite{bonnpr} is excellent. That is not too surprising, 
in both cases the numerical accuracy is determined by the integration over
$\cos \theta'$.   

As second example we take the neutron-proton version of the chiral NNLO potential
\cite{evgeny.report}
and use the same set of the parameters as given in \cite{new2N}.
In Figs.~\ref{f4}--\ref{f6} selected partial wave projected potential
amplitudes are shown for this potential. Again, the agreement between the two
methods is excellent.

\section{Extension to Three-Nucleon Forces}
\label{IV}

The arguments used in Sec.~\ref{II} for the 2N potential can be applied 
to any three-nucleon (3N) force. 
Now we use the 3N states $\mid p q \beta \rangle$ \cite{book}
in the so-called $LS$-coupling
\begin{eqnarray}
\mid p q \beta \rangle \equiv
\mid p q (l \lambda ) L (s \frac12 ) S  (L S ) J M_J \rangle \mid (t \frac12 ) T m_T \rangle  \ , 
\label{pqbeta}
\end{eqnarray}
where the quantum numbers for the relative angular momenta $l$ 
(within the pair $(23)$) and $\lambda$ (between the pair $(23)$ and nucleon $1$)
are coupled to the total angular momentum $L$. In the spin space the spin 
of the  $(23)$ pair is coupled with the spin $\frac12$ of nucleon 1 to the 
total spin $S$. Finally $L$ and $S$ are coupled to the total 3N
angular momentum $J$ with the projection $M_J$. The isospin 3N state,
where we couple the total isospin of the  $(23)$ subsystem $t$ with 
the isospin $\frac12$ of 
the third nucleon to the total 3N isospin $T$ with the projection $m_T$,  
is constructed 
in the same way as the 3N spin state. Since we want the states 
$\mid p q \beta \rangle$ to be antisymmetric 
with respect to the exchange of particles 2 and 3, we require additionally that 
$ \left(-1\right)^{l + s + t} = -1 $. 

We start with the eight-dimensional integral \pagebreak
\begin{eqnarray}
& & \langle p' q' (l' \lambda' ) L' (s' \frac12 ) S'  (L' S' ) J M_J \mid \cr 
& &  \qquad   \qquad  \qquad  \qquad  V^{3N} \mid 
 p q (l \lambda ) L (s \frac12 ) S  (L S ) J M_J  \rangle \nonumber \\
& & = \int d \hat{p\,}' \! \!
\int d \hat{q\,}' \! \!
\int d \hat {p} \! \!
\int d \hat {q}  \nonumber \\
& &  \qquad  \qquad \sum\limits_{m_{L'}} c ( L',S',J;m_{L'}, M_J - m_{L'},M_J) \nonumber \\
& &  \qquad  \qquad \sum\limits_{m_L} c ( L,S,J;m_L, M_J - m_L,M_J)
\nonumber \\
& &  \qquad  \qquad  \qquad 
{\cal Y}_{l',\lambda'}^{*\, L' ,m_{L'}} ({\hat p}',{\hat q}' )  \, 
{\cal Y}_{l ,\lambda }^{ L  ,m_{L }} ({\hat p} ,{\hat q}  )  \, 
\nonumber \\
& & \langle (s \frac12 ) S' M_J -m_{L'}  \mid V^{3N} ( {\fet p'} ,  {\fet q'} , {\fet p} , {\fet q} ) \mid (s \frac12) S  \,M_J - m_L \rangle , \cr
& & 
\label{PWD11}
\end{eqnarray}
where 
\begin{eqnarray}
{\cal Y}_{l, \lambda}^{L, m_L} ( {\hat p} ,{\hat q} ) 
& \equiv & 
\sum\limits_{m_l=-l}^{l} c( l, \lambda, L ; m_l, m_L - m_l , m_L ) \cr 
& & \times Y_{l,m_l} ( {\hat p} ) \, 
Y_{\lambda,m_L-m_l} ( {\hat q} ) .
\label{calY}
\end{eqnarray}
This quantity is independent of $M_J$,  and instead of
Eq.~(\ref{PWD11}) we can calculate
\begin{eqnarray}
& & G ( l',\lambda',L',s',S',l,\lambda,L,s,S,J ) \cr
&  \equiv & \frac{1}{2 J +1} \sum\limits_{M_J=-J}^J \,
\langle p' q' (l' \lambda' ) L' (s' \frac12 ) S'  (L' S' ) J M_J \mid \cr 
& &  \qquad  \qquad  \qquad  \qquad  V^{3N} \mid 
 p q (l \lambda ) L (s \frac12 ) S  (L S ) J M_J \rangle   \ . \cr 
& & 
\label{G}
\end{eqnarray}
Since the integrand is now a scalar, it is possible to reduce
the number of integrals from eight to five. Namely we take first ${\hat p} = {\hat z}$
and then consider all scalar products among the 
${\hat p}$,
${\hat q}$,
${\hat p'}$ and
${\hat q'}$ unit vectors. They depend on the following quantities:
$ \theta_{q}$, 
$ \theta_{p'}$, 
$ \theta_{q'}$, 
$ \phi_{p'} - \phi_{q}$,
$ \phi_{q'} - \phi_{q}$ and
$ \phi_{p'} - \phi_{q'}$. Since the last three are not independent, it is possible to set 
additionally $  \phi_{q}= 0$. Of course, taking  ${\hat p} = {\hat z}$ and $ \phi_{q}= 0$
is only one possibility. The best choice (from the computational point of view)
of the five integration variables 
might depend on the form of 
$ V^{3N} ( {\fet p'} ,  {\fet q'} , {\fet p} , {\fet q} ) $, which is an 
operator in the 3N spin space. 
This quantity $G$ has to be multiplied by the matrix element of isospin operator ${\hat I}$
\begin{eqnarray}
\langle (t' \frac12) T' m_{T'} \mid {\hat I} \mid (t \frac12) T m_{T} \rangle  ,
\label{Iso}
\end{eqnarray}
which can be worked out exactly, independent from the momentum and spin spaces.

\section{Numerical example for a 3N force}
\label{V}

We consider just one example of the 3N force: a two-pion-exchange 
contribution to the chiral NNLO 3N force as given in \cite{chiral3NF},
$V^{3NF}_{TPE}$. 
In particular we take the part of $V^{3NF}_{TPE}$ which is symmetric 
under the exchange of nucleons 2 and 3 \pagebreak
\begin{eqnarray}
V^{(1)}  & = & F_1 
{\fet \sigma_2} \cdot {\fet q_2} \,
{\fet \sigma_3} \cdot {\fet q_3} \,
{\fet \tau_2} \cdot {\fet \tau_3} \ \cr 
& & + \ F_2 
{\fet \sigma_1} \cdot \left(  {\fet q_2}  \times {\fet q_3}  \right) \,
{\fet \sigma_2} \cdot {\fet q_2} \,
{\fet \sigma_3} \cdot {\fet q_3} \,
{\fet \tau_1} \cdot \left(  {\fet \tau_2}  \times  {\fet \tau_3}  \right)  \ , \cr 
& & 
\label{V1}
\end{eqnarray}
where $ {\fet q_i} \equiv {\fet {p'}_i} -{\fet p_i}$
and ${\fet p_i}$ ($ {\fet {p'}_i}$) is the initial (final) momentum 
of nucleon $i$. Further,
$g_A$, $m_\pi$ and $F_\pi$ refer to the nucleon axial vector coupling constant, pion mass
and decay constants, respectively, while $c_i$ are low-energy constants from the
subleading pion-nucleon Lagrangian.
The scalar functions $F_1$ and $F_2$ are 
\begin{eqnarray}
F_1 &  = & \left( \frac{g_A}{2 F_\pi} \right)^2 \,
\frac{1}{
\left(  {\fet q_2}^2 + m_\pi^2 \right) 
\left(  {\fet q_3}^2 + m_\pi^2 \right) 
}
\, \nonumber \\[5pt] 
& & \times  \left( 
- \frac{4 c_1 m_\pi^2 }{F_\pi^2 } + \frac {2 c_3 }{F_\pi^2 } {\fet q_2} \cdot {\fet q_3} \right) 
\end{eqnarray}
and
\begin{eqnarray}
F_2 & = & \left( \frac{g_A}{2 F_\pi} \right)^2 \,
\frac{1}{
\left(  {\fet q_2}^2 + m_\pi^2 \right) 
\left(  {\fet q_3}^2 + m_\pi^2 \right) 
} \, \frac{c_4}{ F_\pi^2 }  \ .
\end{eqnarray}

We consider first the two isospin matrix elements 
\begin{eqnarray}
& & {\hat I}_1 (t',T',m_{T'},t,T,m_T) \cr 
& & \quad \equiv
\langle (t' \frac12) T' m_{T'} \mid {\fet \tau_2} 
\cdot  {\fet \tau_3} \mid (t \frac12) T m_{T} \rangle 
\label{iso1}
\end{eqnarray}
and
\begin{eqnarray}
& & {\hat I}_2 (t',T',m_{T'},t,T,m_T) \cr 
& & \quad \equiv
\langle (t' \frac12) T' m_{T'} \mid {\fet \tau_1} \cdot  \left(
{\fet \tau_2} \times
{\fet \tau_3}
\right)
 \mid (t \frac12) T m_{T} \rangle  .
\label{iso2}
\end{eqnarray}
The matrix elements $ {\hat I}_1 (t',T',m_{T'},t,T,m_T) $
are particularly simple and are given as
\begin{eqnarray}
& & {\hat I}_1 (t',T',m_{T'},t,T,m_T)  \cr 
& & \quad =
\left( 2 t ( t + 1 ) - 3 \right) \,  \delta_{t,t'} \,
 \delta_{T,T'} \, \delta_{m_T,m_{T'}} .
\label{Iso1.2}
\end{eqnarray}
The (purely imaginary) matrix elements  \\ $ {\hat I}_2 (t',T',m_{T'},t,T,m_T) $
can be written as
\begin{eqnarray}
& & {\hat I}_2 (t',T',m_{T'},t,T,m_T) \cr 
& & \quad = 
i \sqrt{3} \, (-1)^{t+1} \,
\delta_{t+t',1} \,
 \delta_{T,\frac12} \, 
 \delta_{T',\frac12} \, 
\delta_{m_T,m_{T'}} .
\label{Iso2.2}
\end{eqnarray}
From Eqs.~(\ref{Iso1.2}) and (\ref{Iso2.2}) we defer immediately that 
the two parts of the considered 3N force (\ref{V1}) will not contribute 
simultaneously to the same matrix element \\
$ G ( l',\lambda',L',s',S',l,\lambda,L,s,S,J ) $.

For this first simple study we construct 16  $\mid \beta \, \rangle $ states 
for $J= \frac12$ and positive parity $\pi = (-1)^{l + \lambda} $ satisfying
the additional condition $l \le 2$ and $\lambda \le 2$. Their quantum numbers 
are given in Table~\ref{tab1}.

\begin{table}
\caption{List of $ \mid \beta \, \rangle $ states for $J^\pi = \frac12^{+}$, 
$l \le 2$ and $\lambda \le 2$.}
\label{tab1}
\begin{center}
\begin{tabular}{ccccccc}
\hline
\quad \quad $\beta$ \quad \quad  &\quad l \quad  & \quad s \quad & \quad $\lambda$ \quad &
\quad L \quad  &\quad S\quad  & \quad t \quad \\
\hline
   1    &  0 & 0 & 0         & 0 & $\frac12$ & 1 \\
   2    &  0 & 1 & 0         & 0 & $\frac12$ & 0 \\
   3    &  0 & 1 & 2         & 2 & $\frac32$ & 0 \\
   4    &  1 & 0 & 1         & 0 & $\frac12$ & 0 \\
   5    &  1 & 0 & 1         & 1 & $\frac12$ & 0 \\
   6    &  1 & 1 & 1         & 0 & $\frac12$ & 1 \\
   7    &  1 & 1 & 1         & 1 & $\frac12$ & 1 \\
   8    &  1 & 1 & 1         & 1 & $\frac32$ & 1 \\
   9    &  1 & 1 & 1         & 2 & $\frac32$ & 1 \\
  10    &  2 & 1 & 0         & 2 & $\frac32$ & 0 \\
  11    &  2 & 0 & 2         & 0 & $\frac12$ & 1 \\
  12    &  2 & 0 & 2         & 1 & $\frac12$ & 1 \\
  13    &  2 & 1 & 2         & 0 & $\frac12$ & 0 \\
  14    &  2 & 1 & 2         & 1 & $\frac12$ & 0 \\
  15    &  2 & 1 & 2         & 1 & $\frac32$ & 0 \\
  16    &  2 & 1 & 2         & 2 & $\frac32$ & 0 \\
\end{tabular}
\end{center}

\end{table}

Next we perform the steps described in Sec.~\ref{IV} and obtain 256 integrands 
$\tilde{G} ( l',\lambda',L',s',S',l,\lambda,L,s,S,J ) \equiv \tilde{G} ( \beta' , \beta \, )$
such that
\begin{eqnarray}
G ( \beta' , \beta \, ) & \equiv &
\langle (t' \frac12) T' m_{T'} \mid 
\langle p' q' \beta' \mid 
V^{3N}
\mid p q \beta \rangle \mid (t \frac12) T m_{T} \rangle \cr 
& \equiv & 
\int d \hat{p\,}' \! \!
\int d \hat{q\,}' \! \!
\int d \hat {p} \! \!
\int d \hat {q} \,
\tilde{G} ( \beta' , \beta \, )  \ .
\label{tildeG}
\end{eqnarray}

Here we show just few (relatively simple) examples. (Note that in the first two cases 
the scalar nature of the $ \tilde{G} ( \beta' , \beta \, )$ functions is clearly visible.)
\begin{eqnarray}
\tilde{G}(1,1) & = &  - \frac{1}{16 \pi^2} \, F_1 \, {\hat I}_1 (1,T',m_{T'},1,T,m_T) \,
{\fet q_2} \cdot {\fet q_3} , \cr 
\label{G1:1} 
\tilde{G}(2,1) & = & - \frac{i}{16 \pi^2 \sqrt{3} } \, F_2 \, {\hat I}_2 (0,T',m_{T'},1,T,m_T) \, \cr 
& & \times \left( \left( {\fet q_2} \cdot {\fet q_3} \right)^2 - {\fet q_2}^2 {\fet q_3}^2  \right) , \cr 
\label{G2:1}
\tilde{G}(5,11) & = & \frac{1}{2 \sqrt{3} } \, F_2 \, {\hat I}_2 (0,T',m_{T'},1,T,m_T) \,
\left( {\fet q_2} \cdot {\fet q_3} \right) \,  \nonumber \\
& & \times {\cal Y}_{2,2}^{0,0} ({\hat p},{\hat q} ) 
      \left( \sqrt{2} \left( q_{4x} -i q_{4y} \right) 
{\cal Y}_{1,1}^{*\, 1,-1} ({\hat p}',{\hat q}' ) \right. \cr 
& &  + 2 q_{4z} {\cal Y}_{1,1}^{*\, 1,0} ({\hat p}',{\hat q}' ) \cr 
& & \left. -  \sqrt{2} \left( q_{4x} + i q_{4y} \right)
{\cal Y}_{1,1}^{*\, 1,1} ({\hat p}',{\hat q}' ) \right) \, , \cr 
\label{G5:11}
\tilde{G}(6,12) & = & \frac{1}{6} F_1 \, {\hat I}_1 (1,T',m_{T'},1,T,m_T) 
  {\cal Y}_{1,1}^{*\, 0,0}({\hat p}',{\hat q}') 
\nonumber \\
& & \left(
\sqrt{2} ( q_{2z} ( q_{3x}+iq_{3y}) \right. \cr 
& & - (q_{2x}+iq_{2y})q_ {3z}) {\cal Y}_{2,2}^{1,-1}(\hat{p},\hat{q})
    \nonumber \\
& & \left.
  +  2i(q_{2y}q_{3x}-q_{2x}q_{3y}) {\cal Y}_{2,2}^{1,0}(\hat{p},\hat{q}) \right. \nonumber \\
& & 
  +  \sqrt{2}(q_{2z} ( q_{3x}-iq_{3y}) \cr 
& & \left.
- (q_{2x}-iq_{2y})q_ {3z}) {\cal Y}_{2,2}^{1,1}(\hat{p},\hat{q})
\right) ,
\label{G6:12}
\end{eqnarray}
where ${\fet q_4} \equiv {\fet q_2}  \times {\fet q_3} $ and 
the Cartesian components of the ${\fet q_i}$ vectors are denoted as 
$q_{ix}$, $q_{iy}$ and $q_{iz}$. Of course, vectors ${\fet q_i}$ 
are now expressed in terms of the initial and final Jacobi momenta 
in the following way
\begin{eqnarray}
{\fet q_1} &=& {\fet q'} - {\fet q} \nonumber \\
{\fet q_2} &=& {\fet p'} - \frac12 {\fet q'} - \left(  {\fet p} - \frac12 {\fet q} \right) \nonumber \\
{\fet q_3} &=& - {\fet p'} - \frac12 {\fet q'} - \left(  -{\fet p} - \frac12 {\fet q} \right) .
\label{qi}
\end{eqnarray}

Many of the $\tilde{G} (\beta' , \beta )$ functions are very lengthy.
Using symbolic manipulations software like {\em Mathematica \copyright} \cite{math}  one can write down these expressions 
directly as a part of a Fortran or a C code.  
The expressions given above have to be integrated over five angles.
With the choice of variables discussed in Sec.~\ref{IV}.
we calculated the five fold integrals for fixed magnitudes of the momenta $p$, $q$, $p'$ and $q'$.
They were chosen quite arbitrarily: 
$p$= 1 fm$^{-1}$,  
$q$= 2 fm$^{-1}$,  
$p'$= 3 fm$^{-1}$,  
$q'$= 4 fm$^{-1}$.
For the isospin quantum numbers we assumed that $T= T' = m_T = m_{T'} = \frac12$.
The 3N force parameters were taken as 
$g_A$= 1.29,
$F_\pi$= 92.4 MeV,
$m_{\pi}$= 138.0 MeV/c$^2$,
$c_1$= -0.81 GeV$^{-1}$,
$c_3$= -3.4 GeV$^{-1}$,
$c_4$= 3.4 GeV$^{-1}$.
Such single five fold integrals can be easily calculated on a PC. 
For a simple test we chose the same number of Gaussian integral points $N$ in each of the five
dimensions and checked the stability of the integrals with respect to $N$. Even in this most simple way, 
we found a good convergence, which is shown
in Table~\ref{tab2}. To check numerically the scalar nature of our integrands
we calculated also the corresponding six fold integrals, where $\phi_{q}$ 
was an additional integration variable and checked their stability
(see Table~\ref{tab3}). We see 
from Tables \ref{tab2} and \ref{tab3}
that the results from the five 
and six fold integrations agree very well with each other. 
We calculated the same matrix elements using standard PWD 
\cite{pwd1,pwd2}. The results are shown in the last row of Table~\ref{tab2}.
They agree very well with the numbers obtained using the new method. 
Thus we conclude that the presented procedure might 
be used to efficiently automate the very cumbersome standard PWD
of the 3N force.
Calculation of all matrix elements 
for the whole grids of $p$, $q$, $p'$ and $q'$ points, especially 
for bigger values of $J$, will require 
an implementation of the algorithm on parallel architectures.
Implementing such a code is much simpler than implementing 
a code based on the traditional PWD. 

\begin{table}
\caption{Stability of the selected five fold integrals with respect
to the number of Gaussian points $N$ (see text). In the last row the 
values obtained with standard PWD \cite{pwd1,pwd2} are given.
All matrix elements are in fm$^{5}$.}
\label{tab2}
\begin{center}
\begin{tabular}{ccccc}
\hline
\quad  $N$ \quad  & \quad $G(1,1)$ \quad  & \quad $G(2,1)$ \quad & \quad $G(6,12)$ \quad & \quad $G(5,11)$ \quad \\
\hline
  12 &  443.565 &  1200.160  & -5.52616 & -5.24720 \\
  24 &  443.618 &  1200.223  & -5.49311 & -5.48527  \\
  36 &  443.618 &  1200.219  & -5.49290 & -5.48630 \\
  48 &  443.618 &  1200.219  & -5.49290 & -5.48626 \\[2pt]
\hline  \\[-8pt]
\parbox{0.8cm}{
standard \\ PWD} &  443.618 &  1200.219  & -5.49274 & -5.48597 \\
\end{tabular}
\end{center}
\end{table}

\begin{table}
\caption{Stability of the selected six fold integrals with respect 
to the number of Gaussian points $N$ (see text).
All matrix elements are in fm$^{5}$.}
\label{tab3}
\begin{center}
\begin{tabular}{ccccc}
\hline
\quad  $N$ \quad  & \quad $G(1,1)$ \quad  & \quad $G(2,1)$ \quad & \quad $G(6,12)$ \quad & \quad $G(5,11)$ \quad \\
\hline
  12 &  443.510 &  1200.365  & -5.55695 & -5.14678 \\
  24 &  443.619 &  1200.218  & -5.49277 & -5.48785 \\
  36 &  443.618 &  1200.219  & -5.49290 & -5.48626 \\
  48 &  443.618 &  1200.219  & -5.49290 & -5.48626 \\
\end{tabular}
\end{center}
\end{table}

\section{Summary and Outlook}
\label{VI}
We propose a new general method of calculating the PWD of any NN potential
which is given in momentum space in terms of operators acting in the 2N 
spin and isospin spaces. Once the expansion coefficients 
in the operator basis are determined, the matrix elements of interest 
are obtained as simple one dimensional integrals, independent from 
the particular form of the NN potential. 
We demonstrate this method with  two examples, the one-boson-exchange 
BonnB potential and a recently derived chiral NNLO potential.
In both cases the agreement between our new method and traditionally employed
one to obtaining partial wave projected matrix elements is perfect.

This method can be extended to treat three-nucleon forces, for which
a traditional PWD is a formidable task.
In order to show the power of our new method, 
we performed a simple feasibility study for the TPE 
chiral 3N force.
A very good agreement between the new method and standard PWD shows
that the very complex traditional approach can be replaced by this straightforward 
and efficient algebraic method. An `automated' PWD for the 3NF is 
of great importance in view of awaiting applications in the 3N systems.
The numerous contributions to the chiral N3LO 3N force 
\cite{Bernard:2007sp,Ishikawa:2007zz} 
can be handled only 
with such an efficient and simple procedure. 

\section*{Acknowledgments}
This work was supported by the Polish 2008-2011 science funds as the
research project No. N N202 077435 and in part under the
auspices of the U.~S.  Department of Energy, Office of
Nuclear Physics under contract No. DE-FG02-93ER40756
with Ohio University.
It was also partially supported by the Helmholtz
Association through funds provided to the virtual institute ``Spin
and strong QCD''(VH-VI-231)
and to the young investigator 
group ``Few-Nucleon Systems in Chiral Effective Field Theory'' 
(grant VH-NG-222)
and by the European Community-Research Infrastructure
Integrating Activity
``Study of Stron\-gly Interacting Matter'' (acronym HadronPhysics2,
Grant A\-gree\-ment n. 227431)
under the Seventh Framework Programme of EU.
 The numerical
calculations have been par\-tly performed on the
 supercomputer cluster of the JSC, J\"ulich, Germany.

\appendix

\section{Integrals for the 2N potential}
\label{A}

This appendix collects the expressions for the partial wave decomposition
of the NN potential for the partial waves with total 2N angular momentum
$j \leq 2$. 
The expressions for higher $j$  can be obtained on 
request from one of the authors (JG) either in analytical form
or as Fortran  or C code. 

\noindent
The second function for $j = 0$ is  ($ H(0,0,0,0)$ was given in Sec.~\ref{II}): \pagebreak
\begin{eqnarray}
& & H(1,1,1,0) =   2 \pi   \int\limits_{-1}^{1} dx \,
\Bigg\{ \left[ f(1) +f(2) \right] x \cr 
& & \quad + 2 f(3)\; p p' (x^2-1)  
   + f(4)\; p^2p'^2 x (1-x^2) \nonumber \\[5pt]
& & \quad - f(5)\;\left[ 2 p'p +x(p'^2 +p^2) \right] \cr 
& & \quad   +  f(6)\; \left[ 2 p'p - x(p'^2 +p^2) \right] \Bigg\}  \ .
\end{eqnarray}
\noindent
The functions for $j=1$ are given by
\begin{eqnarray}
& & H(1,1,0,1) =   2 \pi   \int\limits_{-1}^{1} dx \,
 \Bigg\{ \left[ f(1) + 3 f(2) \right] x \cr 
& & \quad   + f(4)\; p^2p'^2 x (x^2-1) \nonumber \\[5pt]
& & \quad - f(5)\; x \left[ 2 p'p x + (p'^2 +p^2) \right] \cr
& & \quad    +  f(6)\; x \left[ 2 p'p x - (p'^2 +p^2) \right] \Bigg\}  \ , \\
& & H(1,1,1,1) =   2 \pi   \int\limits_{-1}^{1} dx \,
 \Bigg\{ \left[ f(1) +f(2) \right] x \cr 
& & \quad   + f(3) \; p'p (x^2-1) \nonumber \\[5pt]
& & \quad    + f(5)\; \left[ x ( p'^2+p^2) +p'p (x^2+1) \right] \cr 
& & \quad    +  f(6)\; \left[ x ( p'^2+p^2) - p'p (x^2+1) \right] \Bigg\}  \ , \\ 
& & H(0,0,1,1) =  \frac{2\pi}{3} \int\limits_{-1}^{1} dx \,
 \Bigg\{ 3 \left[ f(1) +f(2) \right]  \cr
& & \quad     + f(4)\; p^2p'^2 \; (1-x^2) \cr
& & \quad    + f(5)\; \left[ p'^2 +p^2 + 2p'px \right] \cr 
& & \quad    +  f(6)\; \left[ p'^2 +p^2 - 2p'px \right] \Bigg\}   \ , \\
& & H(0,2,1,1) =  2\pi \frac{\sqrt{2}}{3} \int\limits_{-1}^{1} dx \,
  \Bigg\{ f(4) \; p'^2 p^2 (x^2-1) \cr
& & \quad + f(5)\; \left[ (3x^2-1)p'^2 +2p^2 +4p'px \right] 
      \nonumber \\[5pt] 
& & \quad +  f(6)\;\left[ (3x^2-1)p'^2 +2p^2 -4p'px \right] \Bigg\}  \ , \\
& & H(2,0,1,1) =  2\pi \frac{\sqrt{2}}{3} \int\limits_{-1}^{1} dx \,
 \Bigg\{ f(4) \; p'^2 p^2 (x^2-1) \cr
& & \quad + f(5)\; \left[ 2p'^2 +(x^2-1)p^2 +4p'px \right] 
\nonumber \\[5pt] 
& & \quad  +  f(6)\;\left[ 2p'^2 +(x^2-1)p^2 -4p'px \right]  \Bigg\}  \ , 
\end{eqnarray}
and \pagebreak
\begin{eqnarray}
& & H(2,2,1,1) =  \frac{\pi}{3} \int\limits_{-1}^{1} dx \, 
 \Bigg\{ 3 \left[ f(1) + f(2) \right] (3x^2-1)  \cr 
 & & \quad + 18 f(3)\; p'px(x^2-1)  \nonumber \\[5pt] 
& & \quad + f(4) \; p'^2 p^2 (14x^2 -5 -9x^4) \nonumber \\[5pt] 
& & \quad  + f(5)\; \left[(1-3x^2) (p'^2+p^2) -4p'px \right] \nonumber \\[5pt] 
& & \quad + f(6)\;\left[ (1-3x^2) (p'^2+p^2) +4p'px \right]  \Bigg\}  \ .
\end{eqnarray}

\noindent
The functions for $j=2$ are given by
\begin{eqnarray}
& & H(2,2,0,2) =  \pi \int\limits_{-1}^{1} dx \, (1-3x^2)
 \Bigg\{ \left[-f(1) + 3 f(2) \right] \cr 
 & & \quad - f(4) \; p'^2 p^2 (x^2-1) \nonumber \\[5pt] 
& & \quad + f(5) \; \left[ (p'^2+p^2)  +2 p'px \right] \nonumber \\[5pt] 
& & \quad +  f(6)\;\left[ (p'^2+p^2)  -2 p'px \right] \Bigg\}  \ , \\
& & H(2,2,1,2) =  \pi \int\limits_{-1}^{1} dx \, 
\Bigg\{ \left[ f(1) +f(2) \right] (3x^2-1) \cr 
& & \quad + 2 f(3) \; p'px(x^2-1) \nonumber \\[5pt] 
& & \quad  + f(4) \; p'^2 p^2 \left[ x^4 -2x^2 +1 \right] \nonumber \\[5pt] 
 & & \quad + f(5) \; \left[ (3x^2-1)(p'^2+p^2) + 4 p'p x^3 \right] \nonumber \\[5pt] 
& & \quad  + f(6) \; \left[ x(3x^2-1)(p'^2+p^2) - 4 p'p x^3 \right] 
       \Bigg\}  \ , \\
& & H(1,1 ,1,2) =   \frac{2\pi}{5} \int\limits_{-1}^{1} dx \, 
 \Bigg\{ 5x \left[ f(1) + f(2) \right]  \cr 
 & & \quad + 5 f(3) \; p'p(1-x^2)  \nonumber \\[5pt]  
 & & \quad  + 2 f(4) \; p'^2p^2x (1-x^2) \nonumber \\[5pt] 
& & \quad + f(5) \; \left[ x(p'^2+p^2) +p'p (3x^2-1) \right] 
      \nonumber \\[5pt] 
& & \quad  + f(6) \; \left[ x(p'^2+p^2) - p'p (3x^2-1) \right] \Bigg\}  \ , \\
& & H(1,3,1,2)=  2\pi \frac{\sqrt{6}}{5} \int\limits_{-1}^{1} dx \, 
 \Bigg\{ f(4) \; p'^2 p^2 x(x^2-1) \cr 
&& \quad  + f(5) \; \left[ p'^2 x(5x^2-3) + 2p^2x +2p'p (3x^2-1) \right] \nonumber \\[5pt] 
& & \quad + f(6)\; \left[ p'^2 x(5x^2-3) + 2p^2x -2p'p (3x^2-1) \right] \Bigg\}  \ , \cr & & 
\end{eqnarray}
\pagebreak
\begin{eqnarray}
& & H(3,1,1,2)=  2\pi \frac{\sqrt{6}}{5} \int\limits_{-1}^{1} dx \, 
\Bigg\{ f(4) \; p'^2 p^2 x(x^2-1) \cr 
& & \quad  + f(5) \; \left[ 2p'^2 + p^2 x(5x^2-3)  +2p'p (3x^2-1) \right]
     \nonumber \\[5pt] 
&& \quad + f(6)\; \left[ 2p'^2 + p^2 x(5x^2-3) -2p'p (3x^2-1) \right] \Bigg\}  \ , \cr & & 
\end{eqnarray}
\begin{eqnarray}
& & H(3,3,1,2)=  \frac{\pi}{5}  \int\limits_{-1}^{1} dx \, 
  \Bigg\{  5x (5x^2-3) \left[ f(1)+f(2) \right] \cr 
  && \quad + 10 f(3)\; p'p(1+5x^4 6x^2) \nonumber \\[5pt] 
 && \quad + f(4) \; p'^2 p^2 x (44x^2 -25 x^4 -19) \nonumber \\[5pt] 
 && \quad + f(5) \; \left[ (p'^2+p^2)x(3-5x^2) + 2p'p (1-3x^2)\right]
      \nonumber \\[5pt] 
 && \quad + f(6)\; \left[ (p'^2+p^2)x(3-5x^2) - 2p'p (1-3x^2)\right] \Bigg\}  \ . \cr 
 & & 
\end{eqnarray}



\begin{thebibliography}{99}

\bibitem{charl1} Ch.~Elster, W.~Schadow, A.~Nogga and W.~Gl\"ockle,
Few Body Syst. {\bf 27}, 83 (1999).

\bibitem{hang} H.~Liu, Ch.~Elster and W.~Gl\"ockle,
  Phys. Rev.  C {\bf 72}, 054003 (2005).

\bibitem{imam} I. Fachruddin, Ch. Elster, W. Gl\"ockle,
Phys. Rev. C{\bf 63}, 054003 (2001).

\bibitem{teheran} S.~Bayegan, M.~R.~Hadizadeh and M.~Harzchi,
Phys. Rev. C{\bf 77}, 064005 (2008).

\bibitem{2N3N} W. Gl\"ockle, Ch. Elster, J. Golak, R. Skibi\'nski,
H. Wita{\l}a, H. Kamada, arXiv:0906.0321, Few-Body Syst.
DOI 10.1007/s00601-009-0064-1.

\bibitem{new2N} I. Fachruddin, Ch. Elster, W. Gl\"ockle, J. Golak, 
               D. Rozp{\e}dzik, R. Skibi\'nski,
               K. Topolnicki, H. Wita{\l}a, in preparation.


\bibitem{evgeny.report}
  E.~Epelbaum, H.~W.~Hammer and U.~G.~Meissner,
  arXiv:0811.1338 [nucl-th].
  
  

\bibitem{machl} R. Machleidt, Adv. Nucl. Phys. {\bf 19}, 189 (1989).

\bibitem{bonnpr} R. Machleidt, K.~Holinde, Ch.~Elster,
  Physics Reports {\bf 149}, 1 (1987).



\bibitem{wolfenstein} L. Wolfenstein, Phys. Rev. {\bf 96}, 1654 (1954).

\bibitem{math} Wolfram Research, Inc., {\em Mathematica \copyright}, Version 7.0, Champaign, IL (2008).

\bibitem{book} W. Gl\"ockle, {\em The Quantum Mechanical Few-Body Problem},
Springer-Verlag, Berlin-Heidelberg, (1983).


\bibitem{chiral3NF} E. Epelbaum, A. Nogga, W. Gl\"ockle, H. Kamada, Ulf-G. Mei{\ss}ner and H.~Wita{\l}a,
Phys. Rev. C{\bf 66}, 064001 (2002).

\bibitem{pwd1} S. A. Coon, W. Gl\"ockle,  Phys. Rev. C{\bf 23}, 1790 (1981).

\bibitem{pwd2} D. H\"uber, H. Wita{\l}a, A. Nogga, W. Gl\"ockle, and H. Kamada, 
Few-Body Syst. {\bf 22}, 107 (1997).

\bibitem{Bernard:2007sp}
  V.~Bernard, E.~Epelbaum, H.~Krebs and U.~G.~Meissner,
  Phys.\ Rev.\  C {\bf 77}, 064004 (2008).

\bibitem{Ishikawa:2007zz}
  S.~Ishikawa and M.~R.~Robilotta,
  Phys.\ Rev.\  C {\bf 76}, 014006 (2007).

\end{thebibliography}
\end{document}